# Graded index and randomly oriented core-shell silicon nanowires for broadband and wide angle antireflection

P. Pignalosa[1,2], H. Lee[3], L. Qiao[3], M. Tseng[4], and Y. Yi[2,1,5*]

[1]*New York University, New York, NY*
[2]*City University of New York, SI/GC, New York, NY*
[3]*Beijing University of Science and Technology, Beijing 100083*
[4]*Institute of Electro-Optical Engineering (IEO), NCTU, Taiwan 300*
[5]*Massachusetts Institute of Technology, Cambridge, MA 02139*

Antireflection with broadband and wide angle properties is important for a wide range of applications on photovoltaic cells and display. The SiOx shell layer provides a natural antireflection from air to the Si core absorption layer. In this work, we have demonstrated the random core-shell silicon nanowires with both broadband (from 400nm to 900nm) and wide angle (from normal incidence to 60º) antireflection characteristics within *AM1.5* solar spectrum. The graded index structure from the randomly oriented core-shell (Air/$SiO_x$/Si) nanowires may provide a potential avenue to realize a broadband and wide angle antireflection layer.

*e-mail: yys@alum.mit.edu

For next generation photovoltaic cells, one of the key challenges for further development is how to achieve broadband and wide angle antireflection at the front surface within the *AM1.5* solar spectrum. With the rapid progress of nanotechnology, many nano scale photonic devices as small as 30nm have been realized, which are very promising to achieve manipulation of photons at chip scale and having broad applications in renewable energy (photovoltaic cells, solid state lighting), telecommunications and bio medical field[1]. A number of methods for enhancing optical absorption have been proposed, including the use of dielectric photonic structures[2-18] or plasmonic metallic nanoparticles[19-21]. Silicon nanowires (SiNWs) have attracted extensive interests because of its potential applications in future nanoscale renewable energy devices. Many SiNWs have been fabricated by various techniques, such as the Vapor-Liquid-Solid (VLS) method[22-24], Template-based methods[25-26], lithography[27] and nanoelectrochemistry method[28]. Low reflectance SINWs has also been demonstrated[29-30]. In order to realize the broadband and broad angle antireflection, graded index nanowires with high index Si as core layer, low index $SiO_x$ as cladding layer, and randomness are showing promises[31-32].

In this work, we have demonstrated SiNWs with a *core-shell* structure using VLS method which has potential for large area photovoltaic cell applications. The VLS is a successful method that is applied in SiNWs. The core-shell structure shows good light trapping property at wide range of visible light and broad angle, as the graded index from the core-shell ($Si/SiO_x$) nanowire structure provides natural antireflection characteristics. Present SiNW works on photovoltaic field are largely focused on how to fabricate aligned and ordered nanowire arrays to increase device efficiency[33], although the high costs associated with complicated process makes it difficult to apply in large scale. Here, we introduce a structure that can be produced in a large scale and has a core-shell structure with high index Si as core layer and low index $SiO_x$ as cladding layer, which can be utilized in Si based solar cell as the broadband and broad angle antireflection layer. Further analysis shows the reflectivity is decreased significantly across the broadband from 400nm to 900nm and the wide angle from normal incidence to almost 60º.

We fabricated the SiNWs with a heat evaporation method using a tube furnace. An alumina tube with an opening at one end was located inside a horizontal furnace. Silicon powder placed in a crucible, with a silicon substrate covering the top of crucible, was pushed to the end of the alumina tube. The pressure inside the tube was kept at 8$Pa$ until the temperature reached 900℃, then pumping was stopped to continue heating up to 1200℃, keeping this temperature for 4h. After evaporation, silicon wafer was removed together with crucible until furnace cooling to air temperature. Fig. 1 shows the cross-sectional schematic diagrams of SiNWs core-shell structure. The nanostructure of the SiNWs was investigated by Field Emission Scanning Electron Microscopy (SEM) and High Resolution Transmission Electron Microscopy (HRTEM). Optical properties were measured at room temperature with a UV-3010 Spectrophotometer in the range of 400nm to 900nm. Fig.

2 shows the typical SEM image and HRTEM of core-shell SINWs on the surface of silicon wafer. The morphology of SiNW consists of typical core-shell structure and has been studied in several references[34-37], where laser ablation and evaporation methods were adopted in the heat treatment of the mixture of silicon powder and silica powder, the silicon oxide was assisted as catalyst in the growth of SINW. In our fabrication process, we didn't add any silicon oxide in our reactant; similar core-shell structure was still obtained in the end. During the heating process, the surface of silicon wafer was oxidized by residual oxygen. The resulted silicon oxide doesn't cover whole surface of silicon wafer, due to limited oxygen contention. When the temperature is above 900℃, some silicon powder begins to evaporate as silicon vapor. The exposed silicon surface also evaporates, leaving some pore on the silicon surface. The silicon vapor may be oxidized by the residual oxygen from the ambient air atmosphere. After the holding process ends, the temperature in the furnace is decreased. The degree of sub cooling will provide a driving force for the nuclei of silicon. The nuclei process is accompanied with the decomposition of silicon oxide formed at the high temperature. The decomposition has been studied when discussing the mechanism of the nucleation and growth of SiNWs. $2Si_xO \rightarrow Si_{x-1} + SiO \ (x>1)$, $2SiO \rightarrow Si + SiO_2$ [37]. The decomposition prompts the silicon precipitation from the silica, resulting in the silicon nuclei surrounded by silica.

The core-shell SiNW structure has an optical property as broadband and wide angle antireflection layer. We measured the reflectance of the core-shell nanowire structure at the wavelength range from 400nm to 900nm and the angle from normal incidence to wide angle 60º. The reflection measurement results were illustrated in Fig.3. Fig 3a is the imaging comparison of the two samples (w/o SiNW core-shell structures): the left one is the silicon wafer covered with dense SiNWs, and the right one is original silicon wafer as a reference sample. The reference sample is mirror-like with a shining surface and higher reflectivity. The sample with core-shell SiNWs has a darker surface, demonstrating enhanced absorption due to suppression of reflection from the surface. From this comparison, we can clearly see that the sample with SiNWs absorbed more light compared to the bare silicon wafer. Fig. 3b is the measured reflectance of these samples under normal incidence. In our measurement, all reflected light from the sample which was scattered is collected by the integrating sphere and measured by a photodetector covering the wide solar spectrum wavelength range. The reflectivity from the silicon wafer covered with SiWNs is reduced significantly compared to the reference sample without nanowires.

As the sun moves relative to the solar cell surface, the angular dependence of the antireflection properties can vary significantly, it is strongly desirable to make an antireflection surface independent of incident angle. The angular measurement was also investigated in detail.  Fig.4 shows the results of reflectance at wide angle from normal incidence to almost 60º, Fig.4a is the reflectivity for the reference sample and Fig.4b is for the sample with core shell SiNW structure, which demonstrates that our structure has antireflection properties not only at normal incidence light within the whole visible wavelength range, but also within the wide incident angle from normal incidence to almost 60º.

Randomly poised nanowires greatly increase the incident angle tolerance of the light, increasing the chance of internal reflection inside the nanowire and making incident light form wide angle scattering into the absorption layer. Therefore, the random core shell SiNW structure exhibits broadband and wide angle antireflection within the AM1.5 solar spectrum. The initial results of our work on core shell Si nanowires suggest the future studies on some key parameters, like the diameter, thickness, height, periodicity v.s. randomness, and how we can design and engineer the core shell structure, will be very interesting to explore further. To achieve these goals, in our near future works, we are planning to develop our fabrication technique so that the ordered core-shell nanowire structure can be achieved and the differences between the ordered core-shell nanowires and randomly ordered core-shell nanowires can be compared.

In summary, we have demonstrated silicon nanowires with a *core-shell* structure for potential large area photovoltaic cell applications. The core-shell SiNWs increase the light trapping not only in broad solar spectrum wavelength range but also with relatively large angle independence. The fabrication process is simple, cheap and can be potentially used in large scale manufacturing. The graded index from the core-shell (Si/SiO$_x$) nanowire structure provides a natural antireflection characteristic and provides a potential new avenue to fabricate a broadband and wide angle antireflection layer for next generation photovoltaics and renewable energy applications.

We thank the support from Microsystems Technology Laboratory and Center for Materials Science and Engineering at MIT for the access to the facility and measurement equipments.

**References**

1.  W. D. Li, F. Ding, J. Hu, and S. Y. Chou, *Opt. Exp.*, **19**, 3925 (2011)

2.  M. Berginski, J. Hüpkes, M. Schulte, G. Schöpe, H. Stiebig, B. Rech and M. Wuttig, *J. App. Phys,* **101**, 074903 (2007)

3.  A. V. Shah, M. Vanecek, J. Meier, F. Meillaud, J. Guillet, D. Fischer, C. Droz, X. Niquille, S. Fay, E. Vallat-Sauvain, V. Terrazzoni-Daudrix, J. Bailat, *J. of Non-Crystalline Solids* **338-340**, 639 (2004)

4.  L. Hu and G. Chen, *Nano. Lett.*, 3249 (2007)

5.  S. B. Rim, S. Zhao, S. R. Scully M. D. McGehee, and P. Peumans, *Appl. Phys. Lett.,* **91**, 243501 (2007)

6.  J. Zhao, A. Wang, M. A. Green, and F. Ferrazza, *Appl. Phys. Lett.*, **73**, 1991 (1998)

7.  K. L. Chopra, P. D. Paulson, and V. Dutta, *Prog. Photovolt: Res. Appl.* **12**, 69 (2004)

8.  J. Muller, B. Rech, J. Springer, and M. Vanecek, *Solar Energy* **77**, 917 (2004)

9. J. G. Mutitu, S. Shi, C. Chen, T. Creazzo, A. Barnett, C. Honsberg and D. W. Prather, *Opt. Exp.*, **16**, 15238 (2008)

10. L. Zeng, Y. Yi, C. Hong, J. Liu, X. Duan and L. Kimerling, *Appl. Phys. Lett.* **89**, 111111 (2006)

11. D. Zhou and R. Biswas, *J. Appl. Phys.*, **103**, 093102 (2008)

12. H. Sai, Y. Kanamori, K. Arafune, Y. Ohshita and M. Yamaguchi, *Prog. Photovolt. Res. Appl.* **15**, 415 (2007)

13. J. Springer, B. Rech, W. Reetz, J. Muller, and M. Vanecek, *Solar Energy Materials & Solar Cells* **85**, 1 (2005)

14. J. R. Nagel and M. A. Scarpull, *Opt. Exp.,* **18**, A139 (2010)

15. A. Poruba, A. Fejfar, Z. Remes, J. Springer, M. Vanecek, J. Kocka, J. Meier, P. Torres, and A. Shah, *J. Appl. Phys.,* **88**, 148 (2000)

16. S. Fahr, C. Rockstuhl, and F. Lederer, *Appl. Phys. Lett.*, **92**, 171114 (2008)

17. L. Zeng, P. Bermel, Y. Yi, B. A. Alamariu, K. A. Broderick, J. Liu, C. Hong, X. Duan, J. Joannopoulos, and L. C. Kimerling, *Appl. Phys. Lett.* **93** 221105 (2008)

18. F. Tsai, J. Wang, J. Huang, Y. Kiang, and C. C. Yang, *Opt. Exp.,* **18**, A207 (2010)

19. S. Pillai, K. R. Catchpole, T. Trupke, and M. A. Green, *J. Appl. Phys.* **101**, 093105 (2007)

20. K. R. Catchpole and A. Polman, *Appl. Phys. Lett.*, **93**, 191113 (2008)

21. F. J. Beck, A. Polman, and K. R. Catchpole, *J. Appl. Phys.* **105**, 114310 (2009)

22. Di Gao, Rongrui He, Carlo Carraro, Roger T. Howe, Peidong Yang, and Roya

Maboudian, *J. Am. Chem. Soc*. **127**, 4574-4575 (2005)

23. M. K. Sunkara, S. Sharma, and R. Miranda, *Appl. Phys. Lett*., **79**, 1546-1548 (2001)

24. R. S. Wagner and W. C. Ellis, *Appl. Phys. Lett*. **4**, 89 (1964)

25. I. Lombardi, *Chem. Mater.* **18**, 988-991 (2006)

26. S. Shingubara, O. Okino, Y. Sayama, H. Sakaue, T. Takahagi, *Solid-State Electronics*. **43**, 1143 (1999)

27. K. Peng, *Appl. Phys. Lett*. **90**, 163123 (2007)

28. K. Q. Peng, Y. J. Yan, S. P. Gao and J. Zhu, *Adv. Mater*. **14** (16), 1164 (2002)

29. L. Tsakalakos, J. Balch, J. Fronheiser, M.-Y. Shih, S. F. LeBoeuf, M. Pietrzykowski, P. J. Codella, B. A. Korevaar, O. Sulima, J. Rand, A. Davuluru, and U. Rapolc, *Journal of Nanophotonics*, **1**, 013552 (2007)

30. S. K. Srivastava, D. Kumar, P. K. Singh, M. Kar, V. Kumar, M. Husain, *Solar Energy Materials and Solar Cells*, **94**, 1506 (2010)

31. Y. L. Chiew and K. Y. Cheong, *Physica E: Low-Dimensional Systems and Nanostructures*, **42**, 1338 (2010)

32. F. M. Kolb, H. Hofmeister, R. Scholz, M. Zacharias, U. Gösele, D. D. Ma, and S.-T. Lee, *Journal of The Electrochemical Society*, **151**, G472 (2004)

33. E. Garnett and P. Yang, *Nano. Lett*. **10**, 1082–1087 (2010)

34. Y. F. Zhang, Y. H. Tang, N. Wang, C.S. Lee, I. Bello and S.T. Lee. *Journal of Crystal Growth,* **197**, 136 (1999)

35. W. S. Shi, Y. F. Zheng, N. Wang, C. S. Lee and S. T. Lee, *Appl. Phys. Lett*. **78**

(21), 3304 (2001)

36. Z. W. Pan, Z. R. Dai, L. Xu, S. T. Lee and Z. L. Wang, *J. Phys. Chem. B*, **105**, 2507 (2001)

37. N. Wang, Y. H. Tang, Y. F. Zhang, C. S. Lee, and S. T. Lee, *Phys. Rev. B*. **58**, R16024 (1998)

Light

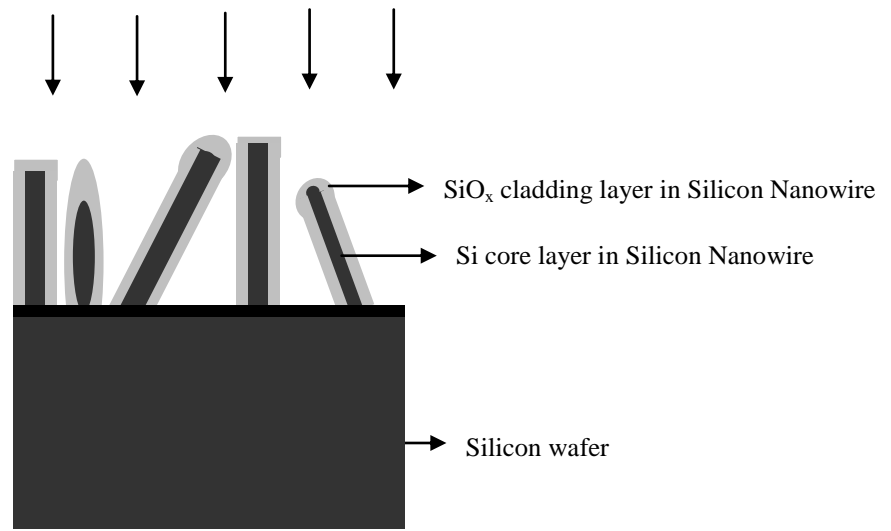

**Figure 1.** Schematic diagrams of core shell nanowires, with high refractive index Si as core layer and low refractive index $SiO_x$ as cladding layer, the whole nanowire structure has a natural graded index layer photonic structure. The core shell nanowires are randomly oriented on the Si surface.

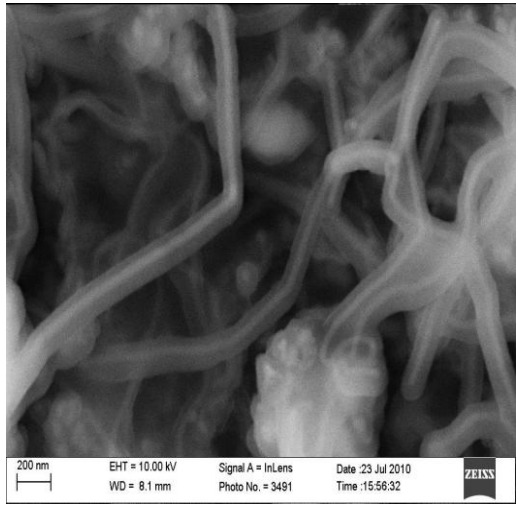

(a)

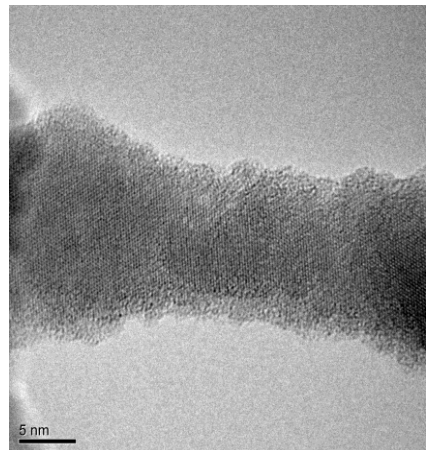

(b)

**Figure 2**. The morphology of core shell silicon nanowires. (a) SEM image shows the typical core-shell structure silicon nanowires. The white color part is silicon nanowire core, while the darker cladding is silicon oxide. (b) HRTEM image (taken from JEM 2010) of a core-shell silicon nanowire structure.

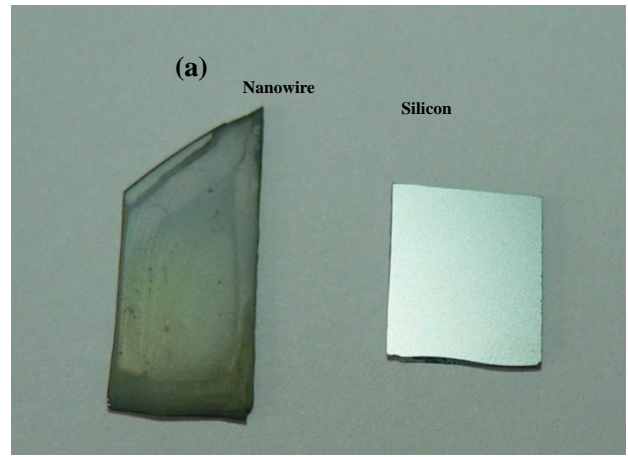

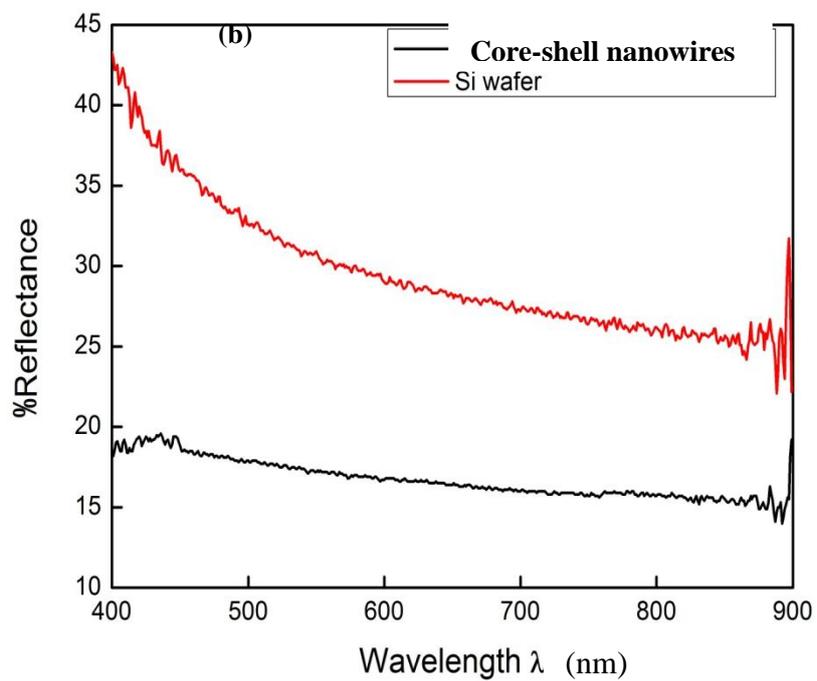

**Figure 3.** (a) Image of silicon wafer covered with core-shell silicon nanowires (left), and original silicon wafer (right). (b) Reflectance of silicon wafer and the silicon wafer covered with core-shell nanowires in the range of 400nm to 900nm. The red line is the reflectance of original silicon wafer from 400nm to 900nm, and the black line corresponds to the reflectance of silicon wafer covered with core-shell nanowires.

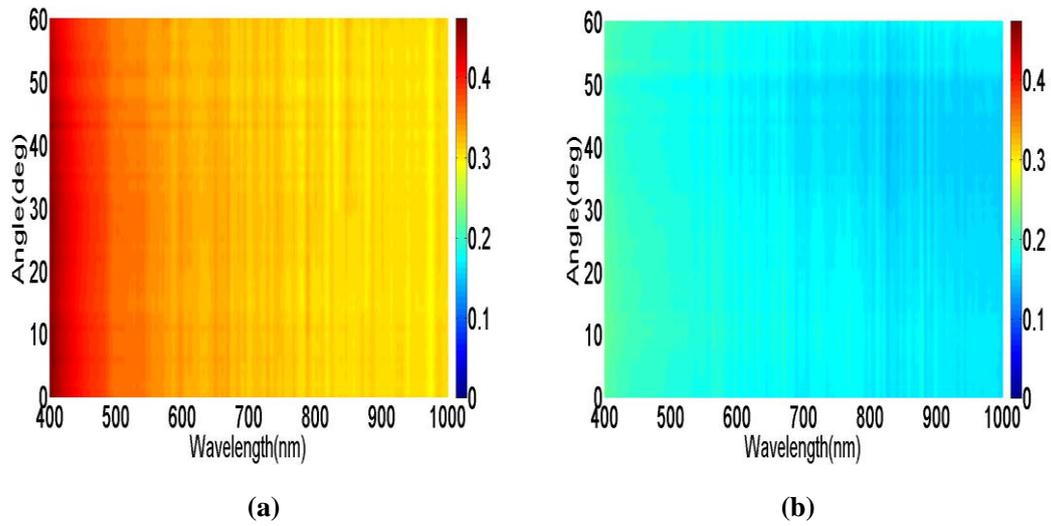

**Figure 4**. Measured results of reflectance on bare silicon wafer (a) and the sample with SiNWs as top layer (b) over wide angle of incidence from normal incidence to 60º. The result shows the lower reflectivity for the sample with core shell SiNWs across broad angle.